\documentstyle[preprint,aps]{revtex}
\begin{document}
\title{Relativistic Dynamics and the Deuteron Axial Current}
\author{B. D. Keister}
\address{
Department of Physics,
Carnegie Mellon University,
Pittsburgh, PA 15213}
\date{\today}
\maketitle
\begin{abstract}
  The deuteron axial current is sensitive both to the form of the
  implementation of relativistic dynamics as well as to the details of
  the deuteron $D$ state at moderate momentum transfer, making it a
  natural partner to the magnetic form factor for exploring details of
  nucleon-nucleon dynamics and associated electroweak properties.
\end{abstract}
\newpage
%
The deuteron has been the testing ground for increasingly refined
theories of the nucleon-nucleon dynamics, from potentials to meson
exchange to quark physics.  It is also one of the simplest venues for
implementing and studying theories which incorporate relativistic
formulations of the dynamics.  Almost all of such formulations have
concentrated on electron scattering, including its form factors
$A(Q^2)$, $B(Q^2)$ and the tensor polarization $T_{20}$.  They include
models inspired by meson-nucleon field theory, such as the
Bethe-Salpeter equation~\cite{Tjon} and the Gross
equation~\cite{Arnold,VanOrden}, and those based upon direct
interactions and light-front dynamics~\cite{LFdeuteron,Karmanov}.  The
results for the electric form factor $A(Q^2)$ indicate rather small
effects from relativity, going beyond a few percent only at momentum
transfers of several GeV${}^2$.  For the magnetic form factor
$B(Q^2)$, calculations based on the Gross equation exhibit sensitivity
to negative-energy $P$-state admixtures~\cite{VanOrden}, and
light-front calculations show marked dependence upon the choice of
matrix elements of $I^+(0)$ used to extract the form
factor~\cite{LFdeuteron}.

It is also important to understand the role of relativistic dynamics
in the deuteron axial current.  This subject was explored in
considerable detail by Frederico, et al., within the framework of
light-front dynamics~\cite{FHM}.  Their primary findings were that (1)
the axial form factor is very sensitive to the choice of matrix
element of $A^+(0)$ -- as much or more so than the magnetic form
factor $B(Q^2)$ -- and (2) this sensitivity is connected almost
entirely to the deuteron $D$ state.  By itself, sensitivity to the
choice of matrix element reflects the fact that the axial current
operator $A^+(0)$ must contain contributions from two-body operators
as a consequence of full rotational covariance at the operator
level~\cite{BKANG}.  The method of imposing covariance implicitly
determines the nature of the two-body currents.  One approach is to
select specific matrix elements for the form factor and to let
covariance determine the others.  This was done explicitly by
Frankfurt, et al., for the deuteron electromagnetic and weak currents
by organizing their matrix elements according to a ``goodness'' vs.\
``badness'' criterion -- a hierarchy which then dictates the choice of
matrix elements for each electroweak form factor~\cite{FFS}.  Another
possible choice is the scheme of Karmanov~\cite{Karmanov}, who sets up
a manifestly covariant framework and extracts amplitudes which do not
depend upon the orientation of the light front in order to achieve
rotational covariance.  This procedure has been applied to matrix
elements of the electromagnetic current.

The purpose of this note is to examine whether the sensitivity to
extraction schemes is merely an artifact of rotational covariance
consistency questions within light-front dynamics, or whether there is
a more general effect which is dependent upon the {\it form} of
relativistic dynamics.  To gain some insight into this question, we
consider the manifestly covariant scheme of Gross~\cite{FG} as a
dynamically distinct alternative to the light-front approach.

The covariant calculation proceeds along the lines described in detail
for electromagnetic currents by Arnold, et al.~\cite{Arnold}.  The
matrix element consists of a momentum loop integral between
deuteron-neutron-proton vertices, with the spectator nucleon
constrained to its mass shell:
\begin{equation}
  \label{Grossmatrixelement}
  G^\mu(Q^2) = \int {d{\bf p}\over2E_{\bf p}(2\pi)^3}
  {\rm Tr}\left[S^T(p) C {\bar\Gamma}^\nu(p',P_d')\xi'_\nu{}^*
    S(P_d' - p) \gamma^\mu\gamma_5
    S(P_d - p) \Gamma^\lambda(p,P_d)\xi_\lambda C\right],
\end{equation}
where $p = (E_{\bf p}, {\bf p})$ is the spectator momentum, $P_d$ and
$P_d'$ the initial and final deuteron momenta, respectively, $E_{\bf p}
= \sqrt{m^2+{\bf p}^2}$, and $S(p) = (\gamma\cdot p -m)^{-1}$.  The
four-vector $G^\mu$ is related to form factors via~\cite{FHM}
\begin{equation}
  \label{FPFA}
  G^\mu(Q^2) = 2P_d^0 {\bf S} F_A(Q^2) - 2{\bf q}({\bf S}\cdot{\bf q})
  \left[F_P(Q^2) + {F_A(Q^2)\over 4(P_d^0+M_d)}\right],
\end{equation}
where ${\bf S}$ is the deuteron spin.  In this work we consider only
$F_A(Q^2)$.  It can be extracted by choosing ${\bf q}$ to lie along
the $z$~axis, and noting that $G^1_{23} = -2iP_d^0F_A$.

The nucleon axial current is taken to be pure $\gamma^\mu\gamma_5$.
One could also supply a nucleon axial form factor which depends upon
$Q^2$ (as well as other variables which describe the extent to which
the struck nucleons are off their mass shells), but for purposes of
comparison these are omitted in the results which are shown, and the
isoscalar nucleon axial coupling constant is set to unity.

Within this scheme, we employ a family of deuteron vertex functions
$\Gamma(p, P_d)$ obtained by Buck and Gross~\cite{BuckGross} for a
range of values of a parameter $\lambda$, which gives the relative
strength of pseudoscalar vs.\ pseudovector coupling ($\lambda=0$ is
pure PV; $\lambda=1$ is pure PS).

For comparison purposes, we include results from a light-front
calculation.  The axial form factor is extracted from matrix elements
\begin{equation}
  \label{Aplus}
  A_{\mu'\mu} := \langle P_d' \mu' | A^+(0) | P_d \mu \rangle.
\end{equation}
Both $A_{10}$ and $A_{11}$ are nonvanishing, and the extracted value
of $F_A(Q^2)$ is quite sensitive to the choice of
$A_{\mu'\mu}$~\cite{FHM}, as noted above.  The results reported here
represent the choice of Frankfurt, et al.~\cite{LFdeuteron}, who
employ a linear combination:
\begin{equation}
  \label{goodcurrent}
  F_A(Q^2) = (1 + \eta)^{-1} (A_{11} + \sqrt{2\eta} A_{10}),
\end{equation}
where $\eta = Q^2 / 4 M_d^2$.

Figure~\ref{FullFig} shows the sensitivity to the form of relativistic
dynamics when all configurations in the deuteron are included.  The
results of the Gross and light-front schemes differ substantially from
each other and from the nonrelativistic calculation.  For $Q^2$ even
as low as 1~GeV${}^2$, there is a noticeable difference among the
calculations.  Figure~\ref{SwaveFig} shows that the contribution to
$F_A(Q^2)$ from the deuteron $S$ state is essentially identical among
the different relativistic formulations as well as the nonrelativistic
limit.  This uniformity was observed for light-front dynamics by
Frederico, et al.~\cite{FHM}, but is evidently also true for the Gross
approach as well.

Part of this sensitivity can be understood from the fact that the
$S$-state contribution to $F_A(Q^2)$ has a node near
$Q^2$=18~fm${}^{-2}$.  When the $D$ state is included, all of the
calculations shown exhibit constructive interference which pushes the
node to higher $Q^2$.  This interference then depends upon the precise
manner in which the $D$-state contribution is implemented.

The interference effect from $S$ and $D$ states suggests that there
should also be a sensitivity to the choice of momentum wave function,
but in fact this is a relatively minor effect.  The light-front
results shown here use wave functions from the Paris~\cite{Paris}
potential.  The same calculation using the Nijmegen~\cite{Nijmegen}
potential gives results where the minimum in $F_A(Q^2)$ moves slightly
(1-2~fm$^{-2}$), but this effect is much smaller than the differences
observed between the forms of relativistic dynamics.

A calculation of current matrix element is not complete without an
accompanying analysis of possible contributions from two-body
currents.  Nonrelativistic calculations can require two-body currents
if the interaction carries charge, as with pion exchange, but
relativistic calculations can require additional two-body
contributions because the current four-vector operator must satisfy
dynamically dependent conditions of relativistic covariance.

One distinctive feature of the Gross equation is that it automatically
includes contributions which manifest themselves as two-body currents
via pair terms ($Z$~graphs) in the nonrelativistic limit.  The
light-front calculations presented here are based on a Hamiltonian
with fixed particle number, rather than a field theory, and therefore
do not automatically contain such terms.  One might then expect that
this difference in content between the two relativistic approaches
explains the quantitative differences shown in the figures.  However,
further investigation reveals that the pair contribution to $F_A(Q^2)$
in the Gross approach is quite small.  Figure~\ref{BGFig} illustrates
several results which should differ significantly from each other if
the physics of pair terms plays an important role.  This can be seen
from the fact that there is little difference among the results for
differing PS/PV ratios $\lambda = 0.0,0.2,0.4$.  Pseudoscalar coupling
gives rise to large pair contributions which then end up as two-body
currents in a calculation which does not include pair excitation.  By
contrast, pseudovector coupling has a separate two-body current
arising from a contact (seagull) interaction, which is not included in
the calculations shown.  Varying $\lambda$ thus illustrates the effect
of the inequivalent treatment of pseudoscalar/pseudovector coupling.
Furthermore, a calculation in which the contribution from the negative
energy $P_s$ and $P_t$ states are omitted, differs little from the
full calculation.  The latter result provides a contrast to the case
of the deuteron magnetic form factor $B(Q^2)$, where the $P$ states
provide important interference effects~\cite{VanOrden}.

In summary, the deuteron axial current $F_A(Q^2)$, together with the
magnetic form factor $B(Q^2)$, provides a sensitive testing ground for
dynamical models, even at moderate $Q^2$.  The effects of relativity
cannot be neglected, and there can be large quantitative differences
among different implementations of relativistic dynamics, specifically
via the deuteron $D$ state.  The manifestly covariant scheme of Gross
exhibits marked sensitivity to $Z$-graph contributions to $B(Q^2)$,
but almost none to $F_A(Q^2)$.  Light-front schemes exhibit a
dependency upon the choice of matrix element used to extract $B(Q^2)$,
and more so to extract $F_A(Q^2)$.

The author wishes to thank Dr.~F. Coester and Prof.~F. Gross for
helpful discussions.  This work was supported in part by the
U.S. National Science Foundation under Grant No.\ PHY-9319641. 
\pagebreak

\begin{figure}
  \caption{Deuteron axial form factor from the Gross
    $\protect\lambda=0.2$ (solid line) and light-front (dashed line)
    relativistic formulations, and the nonrelativistic limit
    (dot-dashed line), using only the full configurations of each
    calculation.}
  \label{FullFig}
\end{figure}

\begin{figure}
  \caption{Deuteron axial form factor from the Gross
    $\protect\lambda=0.2$ (solid line) and light-front (dashed line)
    relativistic formulations, and the nonrelativistic limit
    (dot-dashed line), using only the $S$-wave contribution.}
  \label{SwaveFig}
\end{figure}

\begin{figure}
  \caption{Deuteron axial form factor from the Gross relativistic
    formulation, using Buck-Gross wave functions for
    $\protect\lambda=0.0$ (dashed line), $\protect\lambda=0.2$ (solid
    line), $\protect\lambda=0.4$ (dot-dashed line), and
    $\protect\lambda=0.2$ with $P$ states omitted (dotted line).}
  \label{BGFig}
\end{figure}

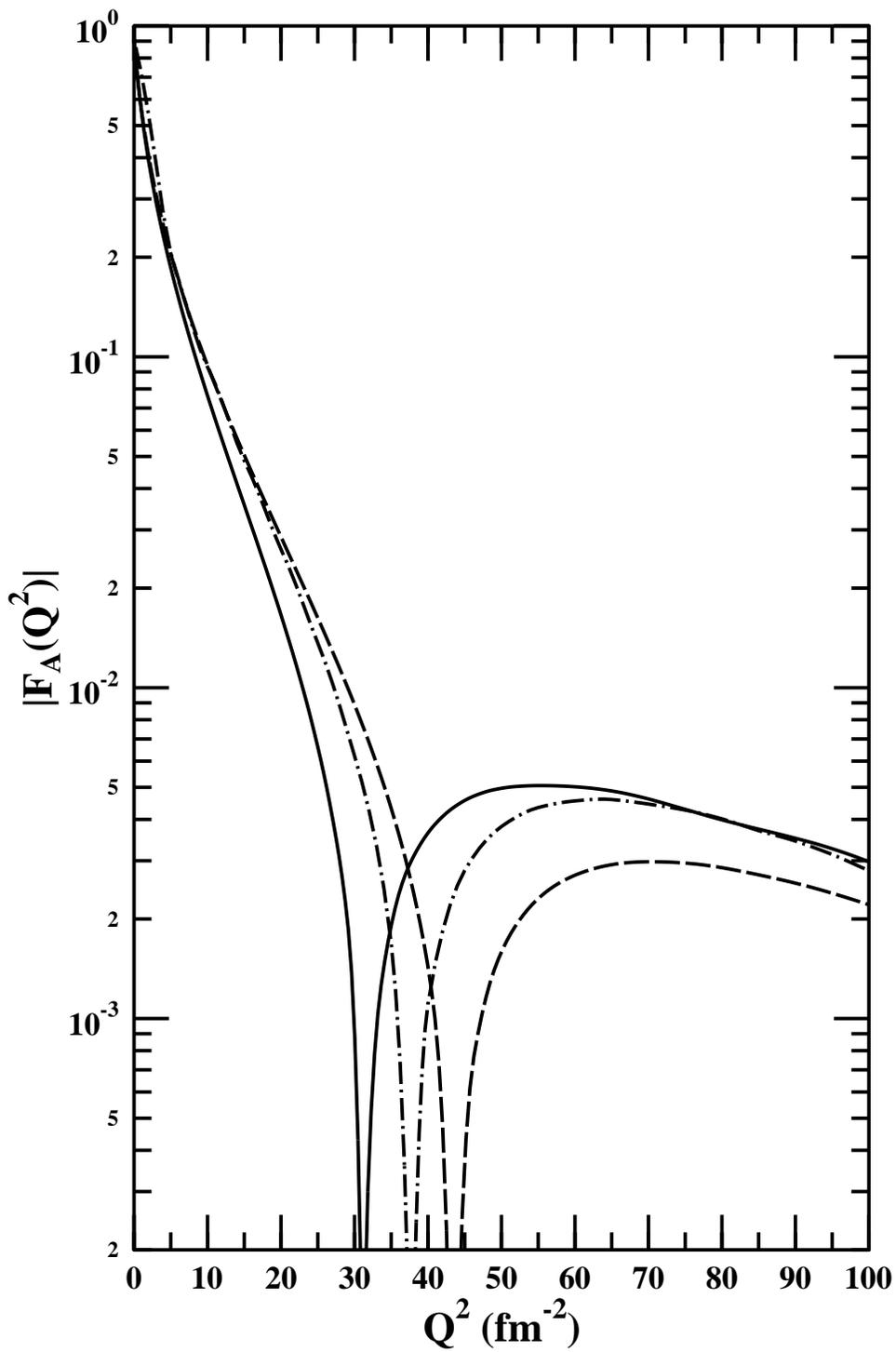
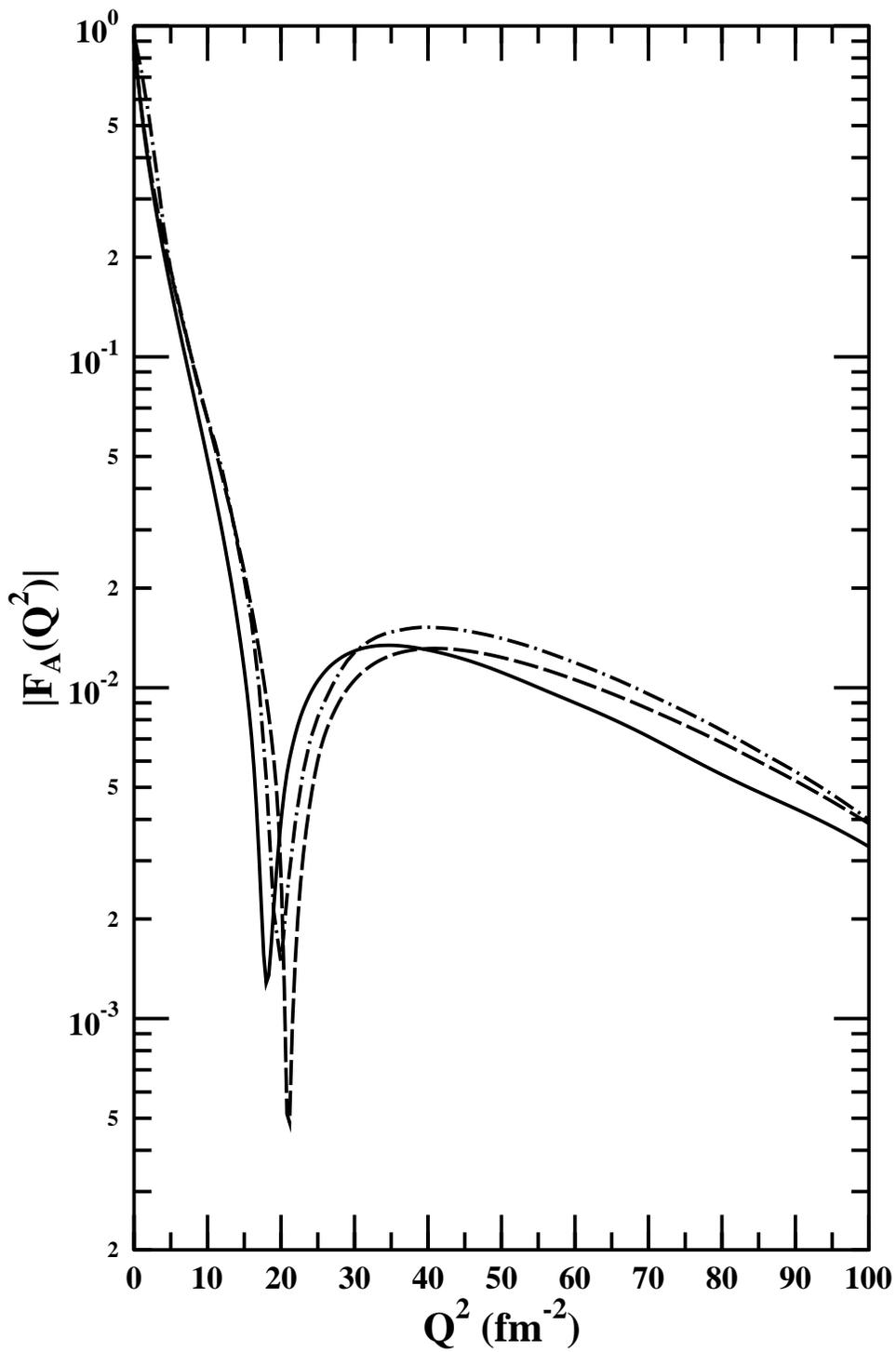
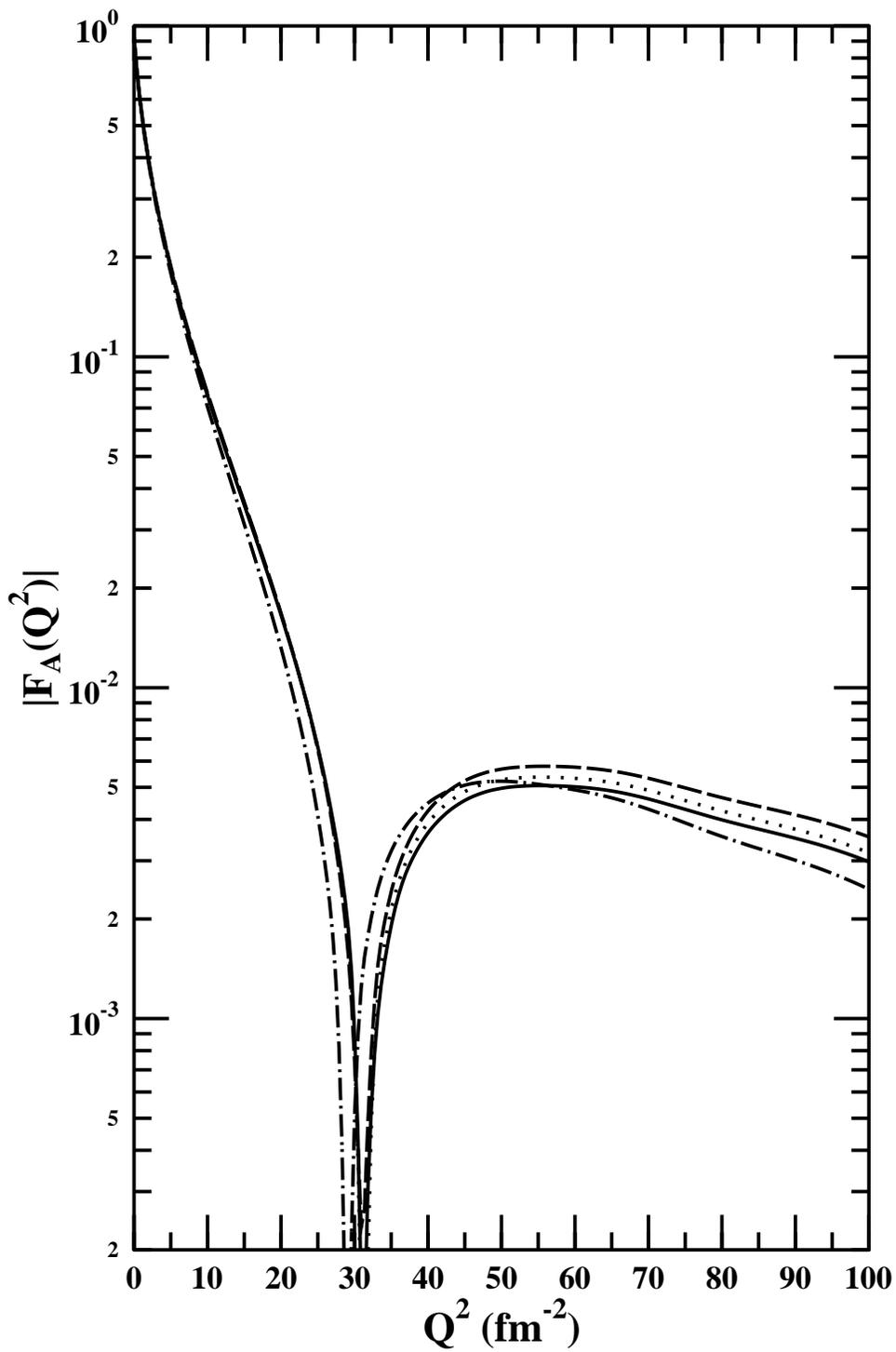
\end{document}